\documentclass[12pt]{article}
\usepackage{graphicx,amsmath}
\usepackage{hyperref}

\newcommand{\beq}{\begin{equation}}
\newcommand{\eeq}{\end{equation}}

\def\*#1{\ensuremath{\vec{#1}}}
\def\+#1{\boldsymbol{#1}}

\title{TRI-BN-25-14: Electrostatic toroidal bender and its fringe fields}
\author{Rick Baartman, TRIUMF}
\date{July, 2025}
\begin{document}
\maketitle
\raggedright

\begin{abstract}
I describe the {\tt COSY} code {\tt GES} that we have used in various forms in the past 30 years to calculate maps to third order through electrostatic bend elements. It has been a mystery that {\tt COSY}'s in-built procedures {\tt ES}, {\tt ESP} and {\tt ECL} disagreed with our own code. This note is intended to clarify the issue.
\end{abstract}

\section{Introduction}
A long time ago (the 90s), I discovered two things concerning fringe fields of optical elements. The first is that the nonlinear effects of `fringe fields' are intrinsic to the elements. That is, they are not reducible by modifying element boundaries. So it was a surprise to me that codes such as {\tt GIOS}\cite{GIOS} and {\tt COSY}\cite{COSY} (and many others) allow higher order (beyond linear) calculations with fringe fields omitted. The cost is that non-specialists tasked with designing (for example) a beamline, design without these effects at the start, thinking that they can deal with their complicating effects later, or not at all. Further, that the elements themselves are somehow at fault, and that there is a way of shaping elements to reduce these effects. 

The second discovery was that the lowest order irreducible effect of the fringe field is independent of the shaping of the fringe field falloff. This is a general result and applies to all beam optical elements\cite{baartman2001eeb}. For solenoids and einzel lenses, this effect is in first order so is correctly handled by all transport codes. But in dipoles, the effect is in second order and for quadrupoles in third order\cite{baartman1997intrinsic,baartman2015quadrupoles}.

I suggest that it would be valuable if the major transport codes include the fringe field effects by default. These maps are simple and would be easy to add. Then the motion would always obey all conservation laws, even with the details of the fringe fields turned off. This is the way in which {\tt TRANSOPTR}\cite{TRI-BN-16-06} operates, but it is only a linear code, so the effects are used only to estimate the rms emittance growth, to be used in optimization calculations. 

In {\tt COSY}, such a second order map of a fringe field would use negligible computing time as compared with Runge-Kutta integrating through an Enge function, but would still give most of the effect needed when for example designing the optics of a transfer line matching section.

In this paper, the theory applied to electrostatic deflectors is recapped, and used to derive the second order `hard-edge' transfer map. A new {\tt COSY} routine named {\tt GES} was written and is described. It agrees with the developed theory (and with {\tt GIOS}). {\tt COSY}'s built in {\tt ESP, ECL} elements agree neither with theory, {\tt GES} nor {\tt GIOS}.

\section{Theory}

To find the transfer map of a fringe field, one must find a canonical transformation that eliminates the terms in the expanded Hamiltonian that become singular in the limit of hard-edge fringe fields. I have done this for quadrupoles and in the particular case of electrostatic bends\cite{baartman2015electrostatic}. But the hard-edge limit is not given explicitly in that reference so will be amplified here.

But first let us recap the theory for the toroidal bend and how the potential function was derived. 

\subsection{Toroidal coordinates}Inside the toroidal bend, there are curvatures in both the $s$ (longitudinal) and the $y$ (off-plane, usually vertical) directions, and these are constant. But in the fringe field regions at entry to and exit from the bend, the former varies from zero to its body constant value, while the $y$ curvature remains constant. The potential is found in a coordinate system corresponding to these curvatures, as an expansion with respect to the reference trajectory, assumed to be at ground potential. In this system, the potential has no $y$-dependence, and is given to third order\footnote{It can be expanded to any order, and the fourth order term is used in the code shown in the Appendix, but not used in the analysis below.} in \cite[eq.\,3]{baartman2015electrostatic}.\begin{equation}\label{pot1}
V_{\rm T}(x,s)=hx-h(k+h)\frac{x^2}{2}+[2h(k^2+kh+h^2)-h'']\frac{x^3}{6}
\end{equation}
$h=h(s)=1/\rho(s)$, and $k=1/\rho_y:=c/\hat{\rho}$, a constant: $c=1$ for spherical electrodes, $c=0$ for cylindrical ones. Note that we assume no $y$-dependence in toroidal coordinates, even in the fringe field region where $h$ is changing.

\subsection{Frenet-Serret Coordinates}Switching to the Frenet-Serret coordinate system, where the $y$ direction has no curvature, we know that there has to be $y$-dependence, and that only even powers of $y$ occur. At the same time, the $x$-dependence for $y=0$ must remain as in the above equation, i.e.\ $V(x,0,s)=V_{\rm T}(x,s)$. Writing the potential in Berz's\cite{COSY} notation,
\beq
V(x,y,s)=\sum_i\sum_ja_{ij}(s)\frac{x^iy^j}{i!j!},
\eeq
we have (remember: $h$ is a function of $s$ but $k$, the vertical curvature, is constant):
\begin{eqnarray}
a_{10}&=&h\\
a_{20}&=&-h(h+k)\\
a_{30}&=&2h(h^2+hk+k^2)-h''.
\end{eqnarray}
Inserting into the Laplace equation with curvature in the $xs$-plane that varies only in the $s$-direction, we can find the remaining coefficients. More simply, we can insert into Berz's recursion relation\cite[eq.\,3.19]{berz1999modern}\footnote{Note that the earlier paper\cite[eq.\,23]{COSY} has a misprint.} to find that $a_{20}+a_{02}+ha_{10}=0$ and $a_{12}+a_{30}+ha_{20}+h^2a_{10}+a_{10}''=0$, giving:
\begin{eqnarray}
a_{02}&=&hk\\
a_{12}&=&-h(h+k)
\end{eqnarray}
In the case of using {\tt COSY}, however, we need not work these out since given $V_{\rm T}(x,s)$ in the plane, it can back out the $y$-dependence using Laplace equation. This is done inside the routine {\tt POFF}.

\subsection{Fringe field effect in the hard edge limit}\label{ffhe}The second order in optics (third in potential) term in the series expanded Hamiltonian that becomes singular in the limit of a hard edge fringe field, is $h''x^3/6$. This results in a shift in $P_x'$ of $h''x^2/2$. Integrating by parts twice for entry into the field where $h$ goes from zero to $1/\hat{\rho}$, we find \beq\label{a|xa} \Delta P_x=-hxx'=-\frac{xP_x}{\hat{\rho}}.\eeq But if $P_x$ is to change discontinuously, $x$ must also change. We can find this from the canonical transformation \cite[eq.\,20]{baartman2015electrostatic}, since we know that while $x$ and $P_x$ suffer shifts in the hard edge limit, the transformed coordinates $X$ and $P_X$ do not. From outside where $h=h'=0$ to inside the field where $h=1/\hat{\rho}$, $h'=0$, we recover the change above in $P_x$, but also \beq\label{x|xx} \Delta x=\frac{x^2}{2\hat{\rho}}.\eeq

These two equations (plus of course the first order identity matrix) are the second order map for entry into the bender field. For exit, the signs are reversed. Note that since the singular coefficient of the $x^3$ term, $h''$, depends only upon the electrode curvature in the bend plane, this second order shift applies to all toroidal cases; cylindrical and spherical bends and anything between. 

\section{{\tt GES}: new {\tt COSY} code}
{\tt GES}, the general electrostatic bender element was originally written (with the help of Dobrin Kaltchev in 1995) as a straightforward application of the general element {\tt GE}, in which the curvature $h$ and the median plane electric potential $V$ are given on a grid of $s$ values. ($s$ is the independent variable, the path length along the reference trajectory.) Crucially, one must supply a function upon which the variation of $h$ and $V$ are based in the region of the fringe fields; in the body of the bender, both are constant. The approach is to simply specify $h(s)$ and derive $V_{\rm T}(x,s)$ from it according to eq.\,\ref{pot1}, above. The fringe field regions use the Enge function already available in {\tt COSY}, as a scaling factor to $h$. But this approach is `kludgy' in the sense that it interpolates using a grid of data even though the analytic formulas for potential and curvature are known. So {\tt GES} has been re-written to supply the functional variation of $h$ and $V$ directly to the equations of motion. A new procedure named {\tt GERB} calls {\tt SDELE} directly with a new element type flag {\tt NSDP:=21}. This is detected in {\tt POTFLD} and calls a new electrostatic bender routine {\tt HVBEND} that calculates $h$ and $V$ directly for use in the Runge-Kutta routine. These new routines are in the Appendix, and the {\tt GES} output maps have been updated in the text below.

{\tt GES} has been written as a general toroidal electrostatic bend. Inputs are, in order, radius/m; angle/degrees; aperture/m (as usual, given as half the $D$ parameter in the Enge mode; kind, the bend-plane radius divided by the non-bend-plane radius of the ground surface e.g.\ 0 for cylindrical, 1 for spherical. 

{\tt GES} makes use of the fringe field mode. In mode 0, it will inform the user that fringe fields are on regardless, and operation is performed as if in mode 1. In this mode, equivalent to ``fringe field integral'' technique, the second order matrices are applied at entrance and exit. This is described below. In mode higher than 1, with aperture of zero, the calculation is as in mode 1. When aperture is $>0$ and mode is $>1$, the full integration is taken through the fringe fields whose shape is given by the usual Enge coefficients (though only the entrance coefficients are used for both entrance and exit).

\section{Results}
\subsection{$45^\circ$ spherical bend}
As a first test for the spherical bend, the code {\tt GES} was run with fringe fields omitted. This gave a transfer map identical (in all 7 displayed digits) to that from using {\tt COSY}'s routine {\tt ESP}, with the setting {\tt FR 0;}.

But with fringe fields, the two codes disagreed in second order. The case shown is where $\hat{\rho}=1$\,metre, aperture radius 1\,cm, bend angle $\pi/4$. For the {\tt ESP} case, the fringe field option {\tt FR 3;} was used.\footnotesize
\begin{verbatim}
GES
  0.7175586    -0.6860108      0.000000      0.000000    -0.7113515     100000
  0.7071458     0.7175586      0.000000      0.000000    -0.2928746     010000
   0.000000      0.000000     0.7071057    -0.7069767      0.000000     001000
   0.000000      0.000000     0.7072391     0.7071057      0.000000     000100
   0.000000      0.000000      0.000000      0.000000      1.000000     000010
  0.2928746     0.7113515      0.000000      0.000000     0.1179272     000001
 -0.3890483    -0.8513432      0.000000      0.000000    -0.7226275     200000
 -0.2080036    -0.7010227      0.000000      0.000000    -0.4555529     110000
 -0.4281918E-01-0.2004865      0.000000      0.000000    -0.3266387     020000
   0.000000      0.000000    -0.2048183     0.7200483      0.000000     101000
   0.000000      0.000000    -0.2073573     0.2964552      0.000000     011000
  0.1062105     0.1523398      0.000000      0.000000    -0.2296059     002000
   0.000000      0.000000      1.214442    -0.2892184      0.000000     100100
   0.000000      0.000000     0.5000050     0.3954617E-02  0.000000     010100
  0.2068684     0.7137543      0.000000      0.000000    -0.4282814E-01 001100
  0.7750413      1.349569      0.000000      0.000000     0.4094747     100001
 -0.1466226     0.6951202      0.000000      0.000000     0.1685871     010001
   0.000000      0.000000     0.2069707     0.3549898      0.000000     001001
 -0.2501038    -0.5017973      0.000000      0.000000    -0.1249593     000200
   0.000000      0.000000    -0.1464292     0.2944885      0.000000     000101
 -0.2499386    -0.1551272      0.000000      0.000000    -0.4948194E-01 000002
 ------------------------------------------------------------------------------
symplectic error: 0.7426237171500225E-011
g2 = 0.2075573046766976E-010; g3 = -.1892827006244602E-009

ESP
  0.7117938    -0.6972614      0.000000      0.000000    -0.7090489     100000
  0.7069680     0.7123671      0.000000      0.000000    -0.2928647     010000
   0.000000      0.000000     0.7118297    -0.6968199      0.000000     001000
   0.000000      0.000000     0.7074332     0.7123143      0.000000     000100
   0.000000      0.000000      0.000000      0.000000      1.000000     000010
  0.2928156     0.7093063      0.000000      0.000000     0.1179428     000001
 -0.4967818    -0.1054895E-01  0.000000      0.000000    -0.3707249     200000
   1.002711    -0.8830090E-02  0.000000      0.000000    -0.7489250     110000
  0.2071271    -0.7090427      0.000000      0.000000    -0.3290377     020000
   0.000000      0.000000     0.1776334E-01-0.9683271      0.000000     101000
   0.000000      0.000000     -1.414055    -0.4050090      0.000000     011000
  0.2140086    -0.6851324      0.000000      0.000000    -0.5779540     002000
   0.000000      0.000000     0.5606538E-03-0.9806706      0.000000     100100
   0.000000      0.000000    -0.1601973E-03  1.012666      0.000000     010100
  -1.006252     0.5536150E-02  0.000000      0.000000     0.2539799     001100
  0.9930520      1.050299      0.000000      0.000000     0.4025517     100001
  0.6111984E-01-0.1054575E-01  0.000000      0.000000     0.1664747     010001
   0.000000      0.000000    -0.9317907E-02 0.6428093      0.000000     001001
 -0.5003768     0.2548899E-02  0.000000      0.000000    -0.1225465     000200
   0.000000      0.000000    -0.3544161     0.9951060      0.000000     000101
 -0.2067942    -0.3589569      0.000000      0.000000    -0.4992080E-01 000002
 ------------------------------------------------------------------------------
symplectic error: 0.9751132434303467E-014


GIOS
 (X,X  )= 7.105065998E-01  (A,X  )=-7.002704594E-01  (T,X  )= 1.804174885E+00
 (X,A  )= 7.071273171E-01  (A,A  )= 7.105065998E-01  (T,A  )= 7.458500488E-01
 (X,G  )= 0.000000000E+00  (A,G  )= 0.000000000E+00  (T,G  )= 5.000000000E-01
 (X,D  )= 2.928945961E-01  (A,D  )= 7.084978343E-01  (T,D  )=-3.005972367E-01

 (X,XX )=-3.969286985E-01  (A,XX )=-8.555639235E-01  (T,XX )= 1.861334945E+00
 (X,XA )=-2.075149663E-01  (A,XA )=-7.045513216E-01  (T,XA )= 1.172127563E+00
 (X,XG )= 0.000000000E+00  (A,XG )= 0.000000000E+00  (T,XG )= 9.020962575E-01
 (X,XD )= 7.870776486E-01  (A,XD )= 1.702892178E+00  (T,XD )=-1.016856405E+00
 (X,AA )=-4.289145704E-02  (A,AA )=-2.045619451E-01  (T,AA )= 8.457223812E-01
 (X,AG )= 0.000000000E+00  (A,AG )= 0.000000000E+00  (T,AG )= 3.729286438E-01
 (X,AD )= 2.071137699E-01  (A,AD )= 7.035775500E-01  (T,AD )=-4.521467154E-02
 (X,GG )= 0.000000000E+00  (A,GG )= 0.000000000E+00  (T,GG )=-1.250000000E-01
 (X,GD )= 0.000000000E+00  (A,GD )= 0.000000000E+00  (T,GD )=-1.503074760E-01
 (X,DD )=-2.499952219E-01  (A,DD )=-5.035219341E-01  (T,DD )= 1.289582712E-01
 (X,YY )= 1.052527701E-01  (A,YY )= 1.498494914E-01  (T,YY )=-2.822076770E-02
 (X,YB )= 2.071097886E-01  (A,YB )= 7.104962826E-01  (T,YB )=-3.911099710E-01
 (X,BB )=-2.500085398E-01  (A,BB )=-5.000025584E-01  (T,BB )= 1.885992853E-01


 (Y,Y  )= 7.071067811E-01  (B,Y  )=-7.070928951E-01
 (Y,B  )= 7.071206674E-01  (B,B  )= 7.071067811E-01

 (Y,YX )=-2.046997404E-01  (B,YX )= 7.143109991E-01
 (Y,YA )=-2.071097886E-01  (B,YA )= 2.953006408E-01
 (Y,YG )= 0.000000000E+00  (B,YG )= 0.000000000E+00
 (Y,YD )= 2.071047476E-01  (B,YD )= 7.080976549E-01
 (Y,BX )= 1.209520330E+00  (B,BX )=-2.904750266E-01
 (Y,BA )= 5.000170797E-01  (B,BA )= 2.399148894E-03
 (Y,BG )= 0.000000000E+00  (B,BG )= 0.000000000E+00
 (Y,BD )= 2.071157580E-01  (B,BD )= 2.938771689E-01
\end{verbatim}\normalsize
(I would like to thank Helmut Weick (GSI) for providing the {\tt GIOS} calculation.)

As one can see, {\tt GES} and {\tt ESP} are nothing alike in second order, but {\tt GES} and {\tt GIOS} are substantially in agreement\footnote{There are 4 that are mysteriously in disagreement; all 4 are second order dispersive: {\tt (X,AD ), (A,XD ), (Y,BD ), (B,YD )}.}; the differences in the few \% range can be explained by the default fringe field integrals that were used in {\tt GIOS}. No effort was made to adjust those or conversely to adjust the Enge coefficients in the {\tt GES} call to improve agreement.

To emphasize the point, and to minimize the effect of the fringe field form, a comparison has also been made with vanishing (actually 0.1\,mm) aperture. Only the horizontal motion lines are shown. Symplectic error as output by {\tt COSY} as well as parameters $g_2$ and $g_3$ from \cite{valetov2020derivation}, which are expected to be zero for symplecticity.

\footnotesize\begin{verbatim}
GES
  0.7071366    -0.7070500      0.000000      0.000000    -0.7071188     100000
  0.7071069     0.7071337      0.000000      0.000000    -0.2928931     010000
 -0.3976300    -0.8537546      0.000000      0.000000    -0.7278071     200000
 -0.2065952    -0.7056108      0.000000      0.000000    -0.4577152     110000
 -0.4337109E-01-0.2061276      0.000000      0.000000    -0.3320926     020000
 symplectic error: 0.1791276000985399E-011, 
 g2= 0.1459810050619126E-010, g3= -.4534211894835494E-010

 ------------------------------------------------------------------------------
ESP
  0.7071544    -0.7070065      0.000000      0.000000    -0.7071262     100000
  0.7071069     0.7071594      0.000000      0.000000    -0.2928931     010000
 -0.4999687    -0.1075439E-03  0.000000      0.000000    -0.3749572     200000
   1.000028    -0.8968466E-04  0.000000      0.000000    -0.7499888     110000
  0.2071069    -0.7071261      0.000000      0.000000    -0.3320790     020000
 symplectic error: 0.2390266367238521E-013,
 g2= 0.2220446049250313E-015, g3= -.5014977148831701E-015
\end{verbatim}

\normalsize
This is for easy comparison with the Valetov paper\cite[sections 4.1.1,4.1.5]{valetov2020derivation}, which makes the claim that the disagreement between {\tt GIOS} and the {\tt ESP} routine in {\tt COSY} is explained by the former being incorrect and the latter correct. In fact I find that the opposite is the case. Note again, that {\tt GES} values for second order agree within about $1\%$ with the {\tt GIOS} values above.

\subsection{Fringe field alone}
As an extreme test, {\tt GES} was run for the entry fringe field alone:\footnotesize
\begin{verbatim}
GES
   1.000000    -0.8522169E-04  0.000000      0.000000    -0.1000001E-03 100000
  0.3319555E-12  1.000000      0.000000      0.000000    -0.4907313E-08 010000
   0.000000      0.000000      1.000000    -0.1000001E-03  0.000000     001000
   0.000000      0.000000     0.3343594E-12  1.000000      0.000000     000100
   0.000000      0.000000      0.000000      0.000000      1.000000     000010
 -0.4907314E-08 0.1000001E-03  0.000000      0.000000    -0.1753337E-12 000001
  0.5000000****-0.5016650E-04  0.000000      0.000000    -0.1139163E-03 200000
  0.2000000E-03 -1.000000****  0.000000      0.000000    -0.1484909E-07 110000
  0.1500000E-07-0.1000000E-03  0.000000      0.000000    -0.5820131E-12 020000
   0.000000      0.000000    -0.1487556E-07 0.1073894E-03  0.000000     101000
   0.000000      0.000000    -0.1001822E-11 0.4875532E-08  0.000000     011000
 -0.2437766E-08 0.5369472E-04  0.000000      0.000000    -0.2500001E-04 002000
   0.000000      0.000000     0.2000003E-03-0.5124498E-08  0.000000     100100
   0.000000      0.000000     0.9814630E-08 0.1980498E-13  0.000000     010100
 -0.1001758E-11 0.1487556E-07  0.000000      0.000000    -0.4907313E-08 001100
  0.2436168E-09 0.1278326E-03  0.000000      0.000000     0.1000001E-03 100001
  0.8548097E-12-0.2436678E-09  0.000000      0.000000     0.9814627E-08 010001
   0.000000      0.000000    -0.4907314E-08 0.5000003E-04  0.000000     001001
  0.4907315E-08-0.1000001E-03  0.000000      0.000000    -0.7515738E-13 000200
   0.000000      0.000000    -0.1506119E-12 0.4907313E-08  0.000000     000101
  0.4907314E-08-0.5000004E-04  0.000000      0.000000     0.5169859E-12 000002
 ------------------------------------------------------------------------------
symplectic error: 0.6266648515380947E-015
g2 = -.2109423746787797E-014 ; g3 = -.5410247948976810E-016
\end{verbatim}\normalsize
Note that this is a nearly unity matrix; the exceptions are the two elements denoted by {\tt ****}. These correspond to $\Delta x/x^2$ and $\Delta P_x/(xP_x)$ and are in agreement with the theoretical values (resp.) \ref{x|xx} and \ref{a|xa}. This confirms that {\tt GES} is consistent with theory of section \ref{ffhe}. 

Further, it allows very fast calculation to second order, using the known entry and exit matrices. Thus, when user specifies a zero aperture in {\tt GES}, the exact second order matrices are used and no integration through the fringe field occurs. However, the maps will only be good to second order and a warning message given if higher order is requested. As is well known, the third and higher orders diverge for zero aperture. 

Usage of third and higher order is still possible maintaining symplecticity in a transport system with a second order electrostatic bender, using {\tt COSY}'s handy symplectifier routine {\tt SY}. From this we discover that eq.\,\ref{a|xa} expands as: \beq P_{x\rm f}=P_x\left(1\mp\frac{x}{\rho}\pm\frac{x^2}{\rho^2}\mp\frac{x^3}{\rho^3}\pm...\right)=\frac{P_x}{1\pm x/\rho}.\eeq However, the validity of higher than second order is questionable, since it depends upon the details of the electric field in the fringe regions, generally not known until after the bender has been designed and the potential function determined with a code such as {\tt OPERA}.

The entrance and exit hard-edge fringe field maps {\tt MAPENT} and {\tt MAPEX} are created in {\tt GES} and used if the aperture is zero. See Appendix for code. If both {\tt COSY - GES} and {\tt GIOS} are run with zero aperture (in {\tt GIOS} case aperture $=10^{-6}$\,m), agreement is to a precision of 5 digits in {\tt GIOS} output.

\subsection{Thin lens limit tests}Whether the fringe field is hard edged or not, the transformed Hamiltonian \cite[eqs.\,21-23]{baartman2015electrostatic} still applies. Let $h=1/\rho$, let $c\equiv k/h$, ({\tt kind} parameter in {\tt GES} input). In a thin lens limit, we find aberration shifts in transverse momenta directly as the nonlinear parts of $P_X'=-\partial H/\partial X$ (similarly for $Y$), integrated through the bend length $L=\rho\theta$. These give: 
\begin{eqnarray}
\Delta P_x&\approx&\frac{\theta}{\rho^2}\left(\left(-1-\frac{3}{\gamma^2}+\left(2+\frac{3}{2\gamma^2}\right)c-c^2\right)x^2+\left(-\frac{c}{2\gamma^2}+c^2\right)y^2\right)\\
\Delta P_y&\approx&\frac{\theta}{\rho^2}\left(-\frac{c}{\gamma^2}+2c^2\right)xy
\end{eqnarray}

Set the bend radius to $\rho=1$\,m again, but angle to $\theta=10$\,mrad so that the length is only 1\,cm and $\theta/\rho^2=0.0100$. 
\subsubsection{Spherical non-relativistic}The extreme non-relativistic ($\gamma=1$) spherical ($c=1$) case would be 
\begin{eqnarray}
\Delta P_x&\approx&0.0100\left(\frac{-3}{2}\,x^2+\frac{1}{2}\,y^2\right)\\
\Delta P_y&\approx&0.0100\,(1\,xy)
\end{eqnarray}
The relevant {\tt COSY} map rows using routine {\tt GES} are:\footnotesize\begin{verbatim}
 -0.7499803E-04-0.1499923E-01  0.000000      0.000000    -0.1249941E-01 200000
   0.000000      0.000000    -0.4999702E-04 0.9999846E-02  0.000000     101000
  0.2499857E-04 0.4999423E-02  0.000000      0.000000    -0.2500081E-02 002000
\end{verbatim}\normalsize
\subsubsection{Spherical extreme relativistic} For extreme relativistic, ($\gamma\gg 1$ not practical for electrostatic, but anyway...)
\begin{eqnarray}
\Delta P_x&\approx&0.0100\left(0\,x^2+1\,y^2\right)\\
\Delta P_y&\approx&0.0100\,(2\,xy)
\end{eqnarray}
The relevant {\tt COSY} map rows using routine {\tt GES} are:\footnotesize\begin{verbatim}
 -0.1807055E-11-0.3614216E-09  0.000000      0.000000    -0.4999224E-02 200000
   0.000000      0.000000    -0.1204678E-11 0.1999967E-01  0.000000     101000
  0.4999833E-04 0.9999333E-02  0.000000      0.000000    -0.3333954E-06 002000
\end{verbatim}\normalsize
\subsubsection{Cylindrical non-relativistic}
 One more comparison non-relativistic cylindrical bend ($c=0$),
\begin{eqnarray}
\Delta P_x&\approx&0.0100\left(-4\,x^2+0\,y^2\right)\\
\Delta P_y&\approx&0.0100\,(0\,xy)
\end{eqnarray}
The relevant {\tt COSY} map row using routine {\tt GES} is:\footnotesize\begin{verbatim}
 -0.1999898E-03-0.3999596E-01  0.000000      0.000000    -0.1499866E-01 200000
\end{verbatim}\normalsize
(the other two relevant rows are missing since they are null).

These three examples all show good agreement with the theory of my 2015 note\cite{baartman2015electrostatic}.

\subsection{A test to show how {\tt GES} and {\tt ESP} differ}
The simple case with non-relativistic spherical bend in the thin lens case was run for both routines, but {\bf without} the $h''(s)x^3/6$ in the {\tt GES} case. See routine {\tt HVBEND} in the Appendix \ref{codes} below: the parameters {\tt hp, hpp} were simply zeroed. The full second order transfer maps are shown below.

\footnotesize{\begin{verbatim}
GES without h''. Symplectic error: 0.3763321858289698E-012
   1.000000    -0.9994499E-02  0.000000      0.000000      0.000000     100000
  0.8320034E-07  1.000000      0.000000      0.000000      0.000000     010000
   0.000000      0.000000      1.000000    -0.9999833E-02  0.000000     001000
   0.000000      0.000000     0.8333369E-07  1.000000      0.000000     000100
 -0.9994332E-04-0.1333500E-05  0.000000      0.000000      0.000000     200000
  0.1999967E-01-0.1108454E-12  0.000000      0.000000      0.000000     110000
  0.4159927E-09-0.9999833E-02  0.000000      0.000000      0.000000     020000
   0.000000      0.000000     0.9993980E-04-0.1998529E-01  0.000000     101000
   0.000000      0.000000    -0.1999967E-01-0.6364327E-09  0.000000     011000
   0.000000      0.000000     0.4996657E-06-0.9994646E-04  0.000000     100100
   0.000000      0.000000     0.1374958E-13 0.1999967E-01  0.000000     010100
  0.9999615E-04-0.9993143E-02  0.000000      0.000000      0.000000     002000
 -0.1999967E-01 0.9994518E-04  0.000000      0.000000      0.000000     001100
 -0.8333359E-09-0.2498287E-06  0.000000      0.000000      0.000000     000200
 ------------------------------------------------------------------------------
ESP. Symplectic error: 0.5662692395342754E-012
   1.000000    -0.9991110E-02  0.000000      0.000000      0.000000     100000
  0.8311424E-07  1.000000      0.000000      0.000000      0.000000     010000
   0.000000      0.000000      1.000000    -0.9991124E-02  0.000000     001000
   0.000000      0.000000     0.8311847E-07  1.000000      0.000000     000100
 -0.9994027E-04-0.7909248E-05  0.000000      0.000000      0.000000     200000
  0.1999967E-01 0.6168177E-07  0.000000      0.000000      0.000000     110000
  0.6900877E-09-0.9999834E-02  0.000000      0.000000      0.000000     020000
   0.000000      0.000000     0.9987366E-04-0.1996859E-01  0.000000     101000
   0.000000      0.000000    -0.1999967E-01-0.2853019E-07  0.000000     011000
   0.000000      0.000000     0.4992473E-06-0.9988032E-04  0.000000     100100
   0.000000      0.000000    -0.3673035E-09 0.1999967E-01  0.000000     010100
  0.9992301E-04-0.9984797E-02  0.000000      0.000000      0.000000     002000
 -0.1999967E-01 0.9994355E-04  0.000000      0.000000      0.000000     001100
 -0.1014843E-08-0.2496177E-06  0.000000      0.000000      0.000000     000200
 ------------------------------------------------------------------------------
\end{verbatim}}\normalsize
The maps are virtually identical. The differences are at the $10^{-4}$ level, and these can be explained by the difference in how the Enge function is used. In the {\tt GES} case, it is applied to $h(s)$, while in {\tt ESP} it is applied to the falloff of the multipole components, so essentially to the potential itself. 

Both calculations are accurately symplectic, as expected, since effectively they each derive from a Hamiltonian.

Somehow the {\tt ESP} routine does not take into account the derivatives of the curvature function $h(s)$ in the fringe regions.

\section{Conclusion}
The two routines {\tt ESP} and {\tt GES} operate in fundamentally different ways. The former calls {\tt POTXZ} to build the potential map in the Cartesian median plane of the bend, while the latter calls {\tt POTS} to build it from info along the reference trajectory in the Frenet-Serret frame. They should agree but I am not sufficiently familiar with {\tt POTXZ} to find why they disagree. I have shown that {\tt ESP} does not have the effect of the curvature derivatives in the fringe field regions.

%\clearpage
\bibliographystyle{elsarticle-num}
\bibliography{/Users/baartman/AllDN/Baartman,/Users/baartman/AllDN/AllDN,/Users/baartman/AllDN/Others}
\clearpage
\appendix \section{Codes}\label{codes}
The general electrostatic toroidal bend code {\tt GES}. 
\footnotesize\begin{verbatim}
-------------------------------------------------------------------------------------
procedure ges r phi aper kind ;
{General Toroidal Electrostatic bend. Good to third order.
AUTHOR:Rick Baartman, TRIUMF.
 Arguments:
  r    = radius/m; 
  phi  = angle/degrees;
  aper = half aperture. Determines FF length according to Enge function. aper=D/2.
  kind = the bend-plane radius divided by the non-bend-plane radius 
         of the ground surface e.g. 0 for Cylindrical, 1 for Spherical}
variable h0 1 ; variable hb 1 ; variable hltot 1 ; variable lenes 1 ; variable mnstep 1 ;
variable I 1 ; variable S1 1 ; variable S2 1 ; variable lenfr 1 ;
VARIABLE MAPENT 4000 8 ; VARIABLE MAPEX 4000 8 ;
hb := 1/r ; h0 := kind*hb ; ppol(1) := hb ; ppol(2) := h0 ; lenes := phi/180*pi*r ;
if lfr<2 ; if lfr<1 ; write 6 'GES always has fringe fields' ; endif ;
   lenfr :=0 ; elseif true ;   
   lenfr := 20*aper ; endif ;
   {total length of FF ; 10 apertures either side = 5D either side.}
ppol(3) := lenfr ;
hltot := (lenfr+lenes)/2 ; ppol(4) := hltot ;
if (lenfr=0.) ; if (no>2)*(lfr>1) ;
   write 6 'WARNING: zero fringe field length calculations have not been validated'
   'for order beyond 2' ;
 endif ;
   LOOP I 1 8 ; MAPENT(I) := XX(i) ; MAPEX(I) := XX(i) ; endloop ;
MAPENT(1) := MAPENT(1)+DD(1)%(-1)/R ; MAPENT(2) := MAPENT(2)-DD(2)%(-1)/R ; sy MAPENT ;
MAPEX(1) := MAPEX(1)-DD(1)%(-1)/R ; MAPEX(2) := MAPEX(2)-DD(2)%(-1)/R ; sy MAPEX ;
   mnstep := lenes/100 ; elseif true ; mnstep := aper/30 ; endif ;

dl -lenfr/2 ; 
if lenfr=0 ; am MAPENT ; endif ;
S1 := -hltot ; S2 := hltot ;
gerb S1 S2 mnstep ;
if lenfr=0 ; am MAPEX ; endif ;
dl -lenfr/2 ; 

endprocedure ;
-------------------------------------------------------------------------------------
\end{verbatim}\normalsize
{\tt GES} calls {\tt GERB}:
\footnotesize\begin{verbatim}
-------------------------------------------------------------------------------------
   PROCEDURE GErb S1 S2 mnstep ; 
      VARIABLE I 1 ; NSDP := 21 ; NPOL := 0 ; 
      LOCSET 0 0 0 0 0 0 ; CE := 'RB' ; LOFF := 2 ;
      SDELE S1 S2 mnstep 10*mnstep (S2-S1) .1 ; UPDATE 1 1 ;
      ENDPROCEDURE ;
-------------------------------------------------------------------------------------
\end{verbatim}\normalsize
As {\tt NSDP>0}, {\tt POTS} calls {\tt HVBEND} with the line \\
{\tt IF NSDP=20 ; INTERP ; ELSEIF NSDP=21 ; HVBEND ; ENDIF ;} \\
replacing \\
{\tt IF NSDP=20 ; INTERP ; ENDIF ;}. \\
Here is {\tt HVBEND}\footnotesize\begin{verbatim}
-------------------------------------------------------------------------------------
      PROCEDURE HVBEND ;
      variable hd 16 ;variable hp 16 ; variable hpp 16 ;
      variable h0 1 ; variable hb 1 ; variable hltot 1 ; variable lenfr 1 ;
	 function form s ; form := enge(1,1,2,-10*s+5) ; endfunction ;
	    function vfun x s3 ;
	    if lenfr>0 ;
	       hd := form((s3+hltot)/lenfr)*form((hltot-s3)/lenfr)*hb ;
	    elseif true ; hd := hb ; endif ;
	    hp := der(2,hd) ;
	    hpp := der(2,hp) ;
	    vfun := hd*x ;
	    vfun := vfun-hd*(h0+hd)*x*x/2 ;
	    vfun := vfun+(2*hd*(h0*h0+h0*hd+hd*hd)-hpp)*x*x*x/6 ;
	    vfun := vfun-(6*hd*(hd+h0)*(hd*hd+h0*h0)
            -4*hp*hp-(7*hd+2*h0)*hpp)*x*x*x*x/24 ;
	    vfun := cons(chie)/1000*vfun ;
	    endfunction ;
      hb := ppol(1) ; h0 := ppol(2) ; lenfr := ppol(3) ; hltot := ppol(4) ;
      V := vfun(0+dd(1),s+dd(2)) ; h := hd ;
      ENDPROCEDURE ;
-------------------------------------------------------------------------------------
\end{verbatim}\normalsize
The routine {\tt HVBEND} lives inside {\tt POTFLD}.

\end{document}